# A COMPARATIVE STUDY OF WESTERN AND CHINESE CLASSICAL MUSIC BASED ON SOUNDSCAPE MODELS

*Jianyu Fan[1], Yi-Hsuan Yang[2], Kui Dong[3], Philippe Pasquier[1]*

[1] Metacreation Lab, Simon Fraser University
[2] Research Center for Information Technology Innovation, Academia Sinica
[3] Department of Music, Dartmouth College
Email: jianyuf@sfu.ca, yang@citi.sinica.edu.tw, pasquier@sfu.ca

**ABSTRACT**

Whether literally or suggestively, the concept of soundscape is alluded in both modern and ancient music. In this study, we examine whether we can analyze and compare Western and Chinese classical music based on soundscape models. We addressed this question through a comparative study. Specifically, corpora of Western classical music excerpts (WCMED) and Chinese classical music excerpts (CCMED) were curated and annotated with emotional valence and arousal through a crowdsourcing experiment. We used a sound event detection (SED) and soundscape emotion recognition (SER) models with transfer learning to predict the perceived emotion of WCMED and CCMED. The results show that both SER and SED models could be used to analyze Chinese and Western classical music. The fact that SER and SED work better on Chinese classical music emotion recognition provides evidence that certain similarities exist between Chinese classical music and soundscape recordings, which permits transferability between machine learning models.

## 1. INTRODUCTION

A soundscape recording is "a recording of the sounds at a given locale at a given time, obtained with one or more fixed or moving microphones" [1]. While sound designers use soundscape recordings for sound design in movies and games, soundscapes are also important for compositions. When studying classical music, we found both Chinese and Western classical music are referring to the concept of soundscape [2-5]. Lüshi Chunqiu, an encyclopedic Chinese classic text compiled around 239 BC, describes songs that were composed based on birds' voices [2]. The book Yue Ji says, "Music is created by nature" [3]. Debussy was inspired by the soundscape in nature for his compositions [4]. Researchers found fractal geometry in bird songs, Bach and Mozart's music [5].

In this study, we aim to investigate whether we can analyze Chinese and Western classical music using deep learning models originally built for soundscape recordings. Typically, such a model is not suitable for music because high-level music information, such as harmony, melody, and orchestration, are not learned by the model. However, since both Chinese and Western classical music literature mention the concept of soundscape [2-5], we intended to test whether such models can analyze Chinese and Western classical music.

A sound event detection (SED) system recognizes sound events in audio signals [6]. Soundscape emotion recognition (SER) is using computational methods for recognizing the perceived emotion of soundscape recordings [7]. We use a pre-trained SED and SER models as feature extractors and use the extracted features as inputs to support vector regression (SVR) models. The goal is to predict the perceived emotion of a corpus of 400 Western classical music excerpts (WCMED) and 400 Chinese classical music excerpts (CCMED). WCMED and CCMED were annotated with perceived valence and arousal via a crowdsourcing method. We examined the accuracy in predicting the perceived emotion of Western and Chinese classical music based on soundscape models.

To further investigate the differences between Western and Chinese classical music, we did a classification task based on a classifier trained on WCMED and CCMED. Moreover, feature analysis is performed to further our understanding of acoustic characteristics of Chinese and Western classical music when compared to soundscape recordings. Our evaluations show the better transferability between soundscape models and CCMED than WCMED. This implies certain level of similarities existing between Chinese classical music and soundscape recordings. The contributions are four-fold:

- We curate WCMED and CCMED, collect emotional annotations through a crowdsourcing study, and share the annotated music dataset. The dataset can be found at http://metacreation.net/ccmed_wcmed_soundscape/.

- We use the combination of a SED model and SVR, and the combination of SER models with SVR to predict the perceived emotion of CCMED and WCMED. The performances of MER of CCMED are significantly better than WCMED.

- We train a classifier based on CCMED and WCMED to classify soundscape recordings, and find most soundscape recordings are classified as Chinese music.

- We analyze loudness, rhythm, tonal and timbre features for the emotion recognition of WCMED, CCMED and soundscape recordings, and find that timbre features work significantly better on both CCMED and soundscape than WCMED.

## 2. RELATED WORKS

### 2.1. Emotion Taxonomy and Model of Emotions

Previous studies discuss two types of emotions are at play when listening to music [8, 9]. Perceived emotions are emotions communicated by the source. For example, the perceived emotion of happy songs is "happy". Induced emotions are reactions that the source provokes in listeners. Namely, induced emotion is what listeners' feel. A listener might feel happy when listening to sad music. We focus on perceived emotions.

There are two families of models of emotion: categorical and dimensional models. Categorical models describe emotions using a set of terms, such as happiness, sadness, fear, anger and disgust. In contrast, dimensional models represent emotions across dimensions in a continuous space. For example, Russell's circumplex model uses a two-dimensional plane: arousal and valence, which represent the level of activation and the level of pleasantness respectively [10]. We use Russell's circumplex model in this study.

### 2.2. Soundscape Emotion Recognition

SER research studies computational systems to recognize the perceived emotion of soundscape recordings. Berglund et al. [11] collected ratings of 116 emotional attributes for soundscape in a survey and did a principal component analysis to select two critical dimensions: pleasantness and eventfulness. Fan et al. [7] designed a system for predicting perceived pleasantness and eventfulness for soundscapes recordings. Later, Fan et al. [12] curated the Emo-soundscapes dataset, which contains annotations of perceived valence and arousal of 1213 soundscape excerpts collected from a crowdsourcing study. They then investigated SER using deep learning approaches [13]. The performance is human competitive (arousal, $R^2$: 0.892; valence, $R^2$: 0.759).

### 2.3. Cross-Cultural Music Emotion Studies

Music emotion recognition (MER) is using computational approaches to recognize the emotions expressed by or perceived from music. Most MER studies have focused on Western music [14, 15]. Few MER studies have compared Chinese and Western music. Wu and Xie experimented with Chinese and Western classical MER by collecting annotations of 20 Western classical music pieces and 20 Chinese classical music pieces and training a Bayesian network for MER [16]. The performance of Western classical MER is better than Chinese classical MER. Researchers have also compared Chinese and Western pop songs. Fan and Casey performed Chinese and UK hit song prediction using 40 weeks of data from pop music charts [17]. They found danceability, energy, tempo and speechiness of UK hits are significantly higher than those of Chinese hits. Later, Yang et al. employed an automatic music tagging system and analyzed tags for the top-50 hits and found Western hits have more diverse genres [18]. Hu and Yang examined the cross-dataset and cross-cultural generalizability of MER models for Western and Chinese Pop songs [19]. They trained MER models with songs in one dataset and tested with those in the other. The results show that within-dataset predictions outperformed cross-dataset predictions in general [19].

## 3. DATASET AND DATA COLLECTION

### 3.1. Corpus Construction

We curated WCMED and CCMED to contain 400 excerpts collected from Western classical music recordings and 400 excerpts collected from Chinese classical music recordings respectively. Our Western classical music recordings are chosen from the Saarland dataset consisting of royalty free audio recordings of 200 pieces and movements from the Western classical music repertoire [20]. The instrumentation is diverse so as to cover various timbres. The instruments performed in these pieces include piano, violin, viola, cello, double-bass, flutes, trumpets, trombone, and xylophone, etc. Excerpts in CCMED are collected from a popular Chinese music-streaming platform [21]. The instruments performed in these pieces include guqin (a plucked seven-string instrument), xun (a globular, vessel flute), pipa (a four-stringed instrument), bianzhong (chime-bells), di (Chinese flute), xiao (a vertical flute), and zheng (Chinese zither), etc.

The duration of each excerpt is 8-20 seconds. This is determined by the recommended duration [14, 22], and balancing the completeness of a phrase and the homogeneity of the timbre of an excerpt. In WCMED, the average duration is 12.41 seconds; the standard deviation is 2.74 seconds. In CCMED, the average duration is 13.07 seconds; the standard deviation is 2.89 seconds. All excerpts are converted to a format in WAV (44,100 Hz sampling frequency, 32 bits precision and mono channel).

### 3.2. Crowdsourcing Experiments and Annotations

We conducted two crowdsourcing experiments to collect emotional annotations (arousal and valence) for WCMED and CCMED individually. Regarding the annotations, we used a ranking-based method. Instead of providing absolute ratings, participants do pairwise comparisons by deciding which audio excerpt has higher arousal/valence. This method simplifies the annotation process, and enhances the inter-annotator reliability [23]. We use a Quicksort algorithm to create pairwise comparisons iteratively. For the first iteration, we select one excerpt as the pivot. All remaining excerpts are to be compared with the pivot. For each comparison, we collect 3 annotations and determine the result to be the one that was provided by at least two annotators. After the first iteration, we end up having two separate sets. We then select a pivot in each subset and create new pairwise comparisons. Eventually, we rank all the excerpts. This method has been used in previous studies [12, 23]. We launched separate crowdsourcing studies for valence and arousal on a platform called Figure Eight [24]. First, annotators read a tutorial, which explains the concept of valence and arousal. Second, annotators take a quiz, which contains 5 gold standard hand-selected comparisons. These pairs are easily comparable regarding valence and arousal. Annotators need to achieve 70% of accuracy in the quiz to continue the task. During the actual annotation task, we tracked their performance by inserting similar gold standard comparisons. Annotators need to provide correct answers to continue the study. An audio excerpt can be played repeatedly. Annotators were required to use headphones during the study. The volume control bar was disabled so that annotators could not adjust the individual volumes. To train regression models, we convert the rankings to ratings. We map the range of ranking values, 1 to 400, to a corresponding rating range of 1.0 to -1.0. This procedure has two assumptions. First, the valence and arousal are in the range of [1.0, -1.0]. Second, the distances between two successive rankings are equal.

A total of 989 annotators from 21 countries participated in the study. We evaluate the Inter-annotator reliability based on percent agreement and Krippendorff's alpha. The percent agreement indicates that annotators agreed on 77.3% (arousal) and 76.1% (valence) of comparisons. The values of Krippendorff's alpha are between 0.21 to 0.40, which indicate a fair level of agreement.

## 4. EXPERIMENTS AND ANALYSIS

### 4.1. Emotion Recognition of WCMED and CCMED based on a Sound Event Detection Model

VGGish [6] is an SED model, which is trained on AudioSet, a large-scale audio dataset containing 2,084,320 human-labeled 10-second audio clips [25]. We used the VGGish model as a feature extractor to convert the audio input into latent feature vectors and fed them as input to SVR models for emotion prediction.

*4.1.1. Raw Feature Extraction and Embedding Extraction*

Given an audio excerpt, the VGGish model computes the log-Mel spectrograms as raw features and then generates 128-D embedding vectors. For each audio file, we remove the first embedding vector and the last embedding vector since the timbre of the beginning and the end of each excerpt are usually different from the middle parts.

**Table 1.** Transfer Learning (SED + SVR) Performances on Arousal and Valence for WCMED and CCMED

| Chinese/Western | Arousal | | Valence | |
|---|---|---|---|---|
| | $R^2$ | MSE | $R^2$ | MSE |
| WCMED | 0.687 | 0.098 | -0.026 | 0.325 |
| CCMED | 0.702 | 0.093 | 0.318 | 0.230 |

*4.1.2. Transfer Learning and Evaluation Results*

SVR is a common kernel method and has been used extensively in MER [14]. We selected the radial basis function kernel and used a grid search method to find the parameters C and gamma.

To evaluate the performance, the initial dataset is shuffled 10 times. Each time, 10% of the dataset is randomly selected for testing, and the remaining 90% is for training. The average result of 10 times validations is used as the final result. We performed the train-test split before the embedding extraction to ensure test samples do not appear in the training set. Because our task focuses on the clip-level prediction rather than the frame-level prediction and we obtained multiple predicted results for one testing sample that contains multiple embedding vectors, we use an ensemble method, which takes the average of the predictions as the final prediction for one testing sample. We used $R^2$ and MSE to indicate performances (Table 1).

*4.1.3. Analysis*

We noticed that the performance of predicting arousal of CCMED is slightly higher than that of WCMED. And the performance of predicting valence of CCMED is significantly higher than that of WCMED. We ran paired t-test for the results ($R^2$) in Table 1. The difference between WCMED and CCMED regarding predicting arousal is not statistically significant. But, there was a significant difference (p<0.001) for valence. In general, the VGGish model works much better on WCMED and CCMED when predicting arousal than predicting valence.

Previous studies indicate arousal is easier to predict [7, 14]. For both soundscape and music, arousal is determined by the level of eventfulness and activation. In contrast, valence of soundscape is mainly determined by the timbre [7] and valence of music is represented by high-level information such as harmony, melody, and orchestration [14]. An SED model does not consider the latter information. The fact that $R^2 < 0$ when predicting valence for WCMED indicates the model is worse than using the average value as the prediction. However, the VGGish model works better on CCMED ($R^2$: 0.318). Moreover, the VGGish model was trained on AudioSet [25], which contains Western classical music but no Chinese classical music. VGGish model works even better on CCMED than WCMED which implies there are certain similarities existing between Chinese classical music and soundscape recordings.

### 4.2. Emotion Recognition of WCMED and CCMED based on Sound Emotion Recognition Models

To train SER models, we used the Emo-soundscape dataset [12], which contains 1213 6-seconds long soundscape clips and rankings of the perceived emotion of 1213 clips in the 2D valence-arousal space. After obtaining pre-trained SER models, we extracted handcrafted features of WCMED and CCMED. Then, we passed them to SER models to extract the embedding vectors. Finally, we used embeddings to train SVR models for predicting the perceived emotion of WCMED and CCMED.

*4.2.1. Feature Extraction and Embedding Extraction*

We used Essentia library [26] to extract 305 dimensions low-level spectrum features, which are mean and standard deviation of frame-level spectrum features. Regarding the feature extraction, we applied a frame size of 2,048 and the hop size of 1,024. Then, we adopt a windowing method to perform data augmentation to artificially enlarge the dataset. For one original feature matrix of one audio excerpt, we selected 80 frames of feature vectors as an augmented data point. Since the sample rate of each excerpt is 44,100 Hz, one augmented data point represents about 1.86 seconds of an audio file. We removed frames at the beginning and the end of the original feature matrix because the timbre of the beginning and the end of each excerpt are usually different from the middle parts. We then did feature normalization for each feature dimension within each corpus.

*4.2.2. Transfer Learning and Evaluation Results*

We trained two SER models using Long Short-Term Memory Recurrent Neural Networks (LSTM-RNN), one for arousal and the other one for valence. The performance of LSTM-RNN for SER has been shown in a previous study [13]. Regarding embedding extraction, we removed the last layer of a pre-trained SER model, which is one neuron with a linear activation function. Then, we input feature matrices to the model and obtained multiple 128-D (number of neurons in the second last layer) embeddings. We used the embedding feature vectors to train SVR models for emotion recognition for WCMED and CCMED. We used the same 10 times shuffle and train-test split as described in Section 4.1.2. The same ensemble method is used. Table 2 shows the performances.

**Table 2.** Transfer Learning (SER + SVR) Performances on Arousal and Valence for WCMED and CCMED

| Chinese/Western | Arousal | | Valence | |
|---|---|---|---|---|
| | $R^2$ | MSE | $R^2$ | MSE |
| WCMED | 0.421 | 0.174 | 0.003 | 0.317 |
| CCMED | 0.485 | 0.157 | 0.153 | 0.269 |

*4.2.3. Analysis*

The results indicate similar findings in Section 4.1.3. The performance of predicting arousal of CCMED is slightly better than WCMED. And the performance of predicting valence of CCMED is significantly higher than that of WCMED. A paired t-test for the results ($R^2$) in Table 2 shows that the difference between WCMED and CCMED regarding predicting arousal is not significant. But, there was a significant difference ($p < 0.01$) in the valence for WCMED and CCMED. Relatively speaking, we found that the overall performance of using SER models is not as good as using the SED model. We think this is because the VGGish model was trained on AudioSet, a large-scale dataset, whereas SER models are trained on Emo-soundscape dataset, which is significantly smaller and less diverse.

### 4.3. Binary Classification Analysis

We trained a binary classifier based on CCMED and WCMED. The goal is to analyze whether soundscape recordings are classified as Chinese classical music or Western classical music. We used the same feature extraction techniques as described in Section 4.2.1 for WCMED, CCMED and Emo-soundscape dataset. Similarity, we ended up having 2,464 feature matrices for Western classical music, 2,609 feature matrices for Chinese classical music, which are combined as the training set. Also, we have 3639 feature matrices for soundscape recordings, which is the testing set. We computed the maximum and minimum of each dimension of feature matrices of WCMED and CCMED and did feature normalization. We trained an LSTM-RNN model. The output layer contains one neuron with a sigmoid activation function. We adopted Adam Optimizer with a learning rate of 0.001. Ten percent of the training set is used as a validation set to select the best mode according to the validation loss within 100 epochs. The binary classification accuracy is 91.93%. Out of 3639 soundscape samples, 3182 samples are classified as Chinese classical music, which is 85.97%. This shows soundscape recordings share more similarities with CCMED than WCMED.

### 4.4. Feature Analysis

To further study the acoustic characteristics of Chinese and Western classical music and soundscape recordings, we conducted a regression analysis with four feature sets on the three datasets. We adopt Essentia to extract features including loudness, rhythm, tonal and timbre. Loudness describes the sensation of intensity of sound. Rhythm depicts the tempo or pulse of a music piece. Tonal features include pitch and harmony. Pitch is the perceptual frequency of sound that can be ordered on a scale from low to high. Harmony refers to the relationship between simultaneously occurring pitches. Timbre refers to the texture of sound that is used to judge whether two sounds having the same loudness and pitch are dissimilar [19]. The number of dimensions of each feature set is shown in the first row of Table 3.

We adopted SVR with the linear kernel to investigate features. The parameter C was optimized by grid searches. We conducted feature selection using the recursive feature elimination (RFE). RFE selects features by recursively removing the least important features from the current feature set until the desired number of features is reached. Moreover, we set the ranging of the number of features from 1 to total dimensions and select the best model.

We examine the performance of using four feature sets separately and together. We did 10-times cross-validation and presented the average performance ($R^2$) of the validation set. The performances are shown in Table 3 and 4. For both arousal and valence, the loudness feature set performs well for all three datasets. This corresponds to the finds in previous studies [7, 19]. Whereas rhythmic features perform better for arousal prediction of WCMED and valence prediction of CCMED, a paired t-test shows it is not significant.

Table 3 shows that pitch and harmony can describe level of intensity and eventfulness of CCMED, but that is not the case for WCMED. According to Table 4, when predicting valence, pitch and harmony are more important for WCMED whereas the texture of the sound is more important for CCMED. Moreover, for valence and arousal prediction, timbre features work significantly better on both CCMED and soundscape than WCMED. This might imply soundscape models extract feature embeddings that better describe tonal and timbre, which are also important for Chinese classical music.

**Table 3.** Performance (in $R^2$) of Feature Sets on Arousal

| Dataset | Loudness (7) | Rhythm (22) | Tonal (14) | Timbre (59) |
|---|---|---|---|---|
| WCMED | 0.553 | **0.395** | 0.046 | 0.691 |
| CCMED | 0.432 | 0.354 | **0.222** | **0.751** |
| Soundscape | **0.806** | 0.356 | 0.122 | 0.737 |

**Table 4.** Performance (in $R^2$) of Feature Sets on Valence

| Dataset | Loudness | Rhythm | Tonal | Timbre |
|---|---|---|---|---|
| WCMED | 0.061 | 0.136 | **0.152** | 0.085 |
| CCMED | 0.127 | 0.162 | 0.070 | 0.283 |
| Soundscape | **0.498** | **0.221** | 0.115 | **0.606** |

## 5. CONCLUSIONS AND FUTURE WORK

This study investigates whether it is effective to analyze Chinese and Western classical music using soundscape models. We found that SED and SER models can analyze music recordings and they performed better on MER tasks for CCMED than WCMED. These results imply certain similarities exist between Chinese classical music and soundscape recordings, the type of similarity that permits transferability between machine learning models. In the future, we are interested in building automatic composition models of Chinese classical music that draw inspirations from soundscapes. We also plan to perform MER on WCMED and CCMED.